\documentclass[twocolumn,showpacs,preprintnumbers,amsmath,amssymb]{revtex4-1}
\usepackage{graphicx}
\usepackage{textcomp}

\begin{document}
\title{Tunable dynamic Fano resonances in coupled-resonant optical waveguide}
  \normalsize
\author{Stefano Longhi}
\address{Dipartimento di Fisica, Politecnico di Milano and Istituto di Fotonica e Nanotecnologie del Consiglio Nazionale delle Ricerche, Piazza L. da Vinci
32, I-20133 Milano, Italy}

%
\bigskip
\begin{abstract}
A route toward lineshape engineering of Fano resonances in photonic structures is theoretically proposed, which uses dynamic modulation of the refractive index of a microcavity. The method is exemplified by considering coupled-resonator optical waveguide systems. An exact Floquet analysis, based on coupled-mode theory, is presented. Two distinct kinds of resonances can be dynamically created, depending on whether the static structure sustains a localized mode or not. In the former case a single Fano resonance arises, which can be tuned in both frequency and line width by varying the refractive index modulation amplitude and frequency. In the latter case two resonances, in the form of narrow asymmetric dips in the transmittance, are found, which can overlap resulting in an electromagnetically-induced transparency effect. 
  \noindent
\end{abstract}


\maketitle

\section{introduction}
Fano resonances \cite{r1}, i.e. asymmetric and sharp line-shapes that result from the interference
of discrete resonance states with broadband continuum states, are ubiquitous in several areas of physics \cite{r2}. In optics, 
they are observed in
systems ranging from waveguide-cavity structures to
plasmonics and metamaterials \cite {r2,r3,r4,r5}, 
promising applications for a wide range of photonic devices such as optical filters, switches, modulators and sensors \cite{r6,r7,r8,r9,r10} .
To this regard, line shape engineering, i.e. the possibility to control and tune frequency, shape and width of the Fano resonance, is of major importance. 
Conventionally, the Fano line shape can
be tuned over wide spectral ranges by carefully altering the geometry of a nanostructure \cite{r11,r12,r13}. However, 
dynamical and fine control of Fano resonances is a highly desirable functionality, which can not be accomplished by material or geometric engineering. So far, a few methods have been suggested for dynamical 
line shape engineering, based on two-beam interference \cite{r14}, phase engineering of the excitation beam \cite{r15},  
the use of hybrid gratings \cite{r16}, and the application of  mechanical stress \cite{r17}.\par In this work a different route to create, shape and fine tuning Fano resonances in photonic 
structures  is theoretically proposed, which is based on dynamic modulation of the resonance frequency of a microcavity. We consider specifically 
dynamic Fano resonances in coupled-resonator optical waveguide (CROW) systems \cite{r18,r19,r20}, which enable an exact analysis based on coupled-mode theory.
The CROW  consists of a homogeneous chain of resonators in which light propagates by virtue of the coupling between adjacent resonators; a Fano resonance is dynamically created by periodic modulation of the resonance frequency of one resonator in the chain. Creation of high-$Q$ resonances by dynamic refractive index modulations based on photonic transitions was previously proposed in Ref.\cite{r21}, however in that system the resonances were not of Fano type.
CROWs have been explored in a variety of material platforms and
resonator types, including photonic-crystal defect cavities, microspheres, microdisks,
and microring resonators \cite{r22,r22bis}, with applications to  light slowing down and storage \cite{r23,r24,r25}, light capturing \cite{r26}, time reversal \cite{r27,r27bis}, sensing \cite{r28}, and 
for the realization of topologically-protected edge states. \cite{r29}
While in static CROW structures Fano resonances are usually realized by side-coupled microcavities \cite{r2,r30}, in our setup they arise dynamically from the interference of different Floquet channels \cite{r31,r32,r33,r33bis,r33tris,r33quatris}, and can be thus controlled by tuning the frequency and amplitude of the modulation signal. 

\section{Photonic structure and Floquet analysis}
Figure 1(a) shows a schematic of the CROW structure, which comprises a chain of coupled micro/nano resonators with the same resonance frequency $\omega_0$ and coupling constant $\kappa$. The resonators can be realized, for example, using photonic crystal defect cavities, coupled microdisk resonators or silicon microrings. The resonance frequency of the resonator at site $n=0$ is assumed to be biased and periodically modulated in time. This can be accomplished by modulation of the microcavity refractive index using various physical mechanisms such as free-carrier-plasma dispersion and electro-optic effects \cite{r34,r35,r36,r37}. In CROW systems, light transport can be described by coupled-mode theory \cite{r18,r24,r26,r27}, which reproduces with excellent accuracy the results obtained by FDTD numerical simulations \cite{r21,r24,r26}. Neglecting resonator losses,  the coupled mode equations for the complex field amplitudes $a_n(t)$ in each cavity read \cite{r26}
\begin{equation}
i \frac{d a_n}{dt}= \omega_0 a_n +\kappa(a_{n+1}+a_{n-1})+\delta_{n,0} \Delta \omega_0(t) a_n
\end{equation}
where $\Delta \omega_0(t)= \sigma+ \Gamma \cos (\Omega t)$ is the resonance shift of the resonator at site $n=0$, which comprises a bias (static) term $\sigma$ and a sinusoidal term of amplitude $\Gamma$ and frequency $\Omega$. For $\Delta \omega_0=0$, the static uniform chain of coupled resonators sustains Bloch states of the form $a_n(t)= \exp[-iqn-i \omega(q)t]$ with the dispersion relation $\omega(q)=\omega_0+2 \kappa \cos q$, where $-\pi \leq q < \pi$ is the Bloch wave number [see Fig.1(b)]. Forward-propagating waves in the chain, corresponding to a positive group velocity $v_g= -(\partial \omega / \partial q)=2 \kappa \sin q$, have a Bloch wave number $q$ in the range $(0, \pi)$, whereas backward propagating waves correspond to $-\pi < q <0$. For $\Delta \omega_0 \neq 0$,  the resonator at site $n=0$ acts as a scattering centre, enabling to reflect  light waves propagating along the CROW at some spectral frequencies inside the CROW transmission band $(\omega_0-2 \kappa, \omega_0+2 \kappa)$.
The high-frequency modulation regime $\Omega \gg \kappa$ simply corresponds to a re-normalization of the coupling constant (see, e.g., \cite{r38}); this case it is not of interest for the onset of dynamic Fano resonances. Here we assume that $\Omega$ and $\Gamma$ are  of the same order of magnitude than the coupling constant $\kappa$. In this regime 
dynamic Fano resonances can arise because light can hop across the modulated resonator following different Floquet paths \cite{r31}. Assuming that a forward-propagating light wave with Bloch wave number $q$ ($ 0<q< \pi$) and frequency $\omega(q)=\omega_0+2 \kappa \cos q$ is incident onto the modulated resonator from the left to the right side of the chain, according to Floquet theory \cite{r31} the exact scattered solution to Eq.(1) has the form 
\begin{widetext}
\begin{equation}
a_n(t) \left \{ 
\begin{array}{cc}
\sum_{\alpha= -\infty}^{\infty}  \left\{ \delta_{\alpha,0} \exp[-iq_{\alpha} (n+1)] + r_{\alpha}(q)   \exp[iq_{\alpha} (n+1)] \right\} \exp(-i \Omega_{\alpha} t) & n \leq -1 \\
\sum_{\alpha= -\infty}^{\infty}  B_{\alpha} \exp(-i \Omega_{\alpha} t) & n=0 \\
\sum_{\alpha= -\infty}^{\infty}  t_{\alpha}(q)  \exp[-iq_{\alpha} (n-1)] \times  \exp(-i \Omega_{\alpha} t) & n \geq 1
\end{array}
\right.
 \end{equation}
\end{widetext}
where $\Omega_{\alpha}= \omega (q) + \alpha \Omega$, $r_{\alpha}(q)$ and $t_{\alpha}(q)$ are the reflection and transmission amplitudes of the various Floquet (scattered) orders $ \alpha=0, \pm 1, \pm2, \pm 3, ...$, $B_{\alpha}(t)$ are the harmonic amplitudes of the field in the modulated resonator at $n=0$, and $q_{\alpha}$ are defined from the relation 
\begin{equation}
\cos q_{\alpha}=\cos q+ \alpha \frac{ \Omega}{ 2 \kappa},
\end{equation}
 with $0 \leq q_{\alpha} \leq \pi$ if $q_{\alpha}$ is real (propagative modes) and ${\rm Im}(q_{\alpha}) <0$ if $q_{\alpha}$ is complex (evanescent modes). The power transmittance $T(q)$ and reflectance $R(q)$ of the modulated resonator can be then calculated as
\begin{equation}
T(q)= \sum_{\langle \alpha \rangle} \frac{v_{g \alpha}}{v_{g 0}} |t_{\alpha}|^2 \;, \;\;\; R(q)= \sum_{\langle \alpha \rangle} \frac{v_{g \alpha}}{v_{g 0}} |r_{\alpha}|^2
\end{equation}
where $v_{g \alpha}= 2 \kappa \sin q_{\alpha}$ is the group velocity at the Bloch wave number $q_{\alpha}$ and the symbol $ \langle ... \rangle$ means that the sum is extended over the indices $\alpha$ corresponding to propagative modes (i.e. $q_{\alpha}$ real).  The terms with $\alpha=0$ in the sums of Eq.(4), i.e. $T_0 \equiv |t_0|^2$ and $R_0 \equiv |r_0|^2$, correspond to elastic scattering, i.e. the frequency of transmitted and reflected photon is not altered by the oscillating potential at site $n=0$. The other terms that contribute to the total transmittance and reflectance, namely $T_{\alpha}(q) \equiv (v_{g \alpha}/v_{g 0}) |t_{\alpha}|^2 $ and $R_{\alpha}(q) \equiv (v_{g \alpha}/v_{g 0}) |r_{\alpha}|^2 $, correspond to inelastic scattering channels, with transmitted and reflected photons with a higher (for $\alpha >0$) or lower (for $\alpha <0$)  energy amount $ \alpha \hbar \Omega$. 
 For a lossless system one has $R+T=1$ and the system is reciprocal, i.e. transmittance is independent of the incidence side.  Substitution of the Ansatz (2) into Eq.(1) and eliminating $r_{\alpha}$ and $B_{\alpha}$ yields the following difference equation for the transmission amplitudes $t_{\alpha}$ of various Floquet orders
\begin{eqnarray}
t_{\alpha} \left(   1-\exp(2iq_{\alpha})+\frac{\sigma}{\kappa} \exp(iq_{\alpha})  \right) \nonumber \\
+ \frac{\Gamma}{2 \kappa} \left(  t_{\alpha+1} \exp(i q_{\alpha+1})  +  t_{\alpha-1} \exp(i q_{\alpha-1}) \right) \\
= \delta_{\alpha,0} \left(  \exp(-2 i q_{\alpha})-1 \right) . \nonumber
\end{eqnarray} 
\section{Dynamic Fano resonances}
Let us first notice that, in the absence of the ac modulation, i.e. for $\Gamma=0$, the solution to Eqs.(5) is given by $t_{\alpha}(q)=\delta_{\alpha,0} t_{st}(q)$, where 
$t_{st}(q)$ is the transmission amplitude of the static CROW, given by
\begin{equation}
t_{st}(q)= \frac{\exp(-2iq)-1}{1-\exp(2iq)+(\sigma / \kappa) \exp(iq)}.
\end{equation}
\begin{figure*}
\includegraphics[scale=0.8]{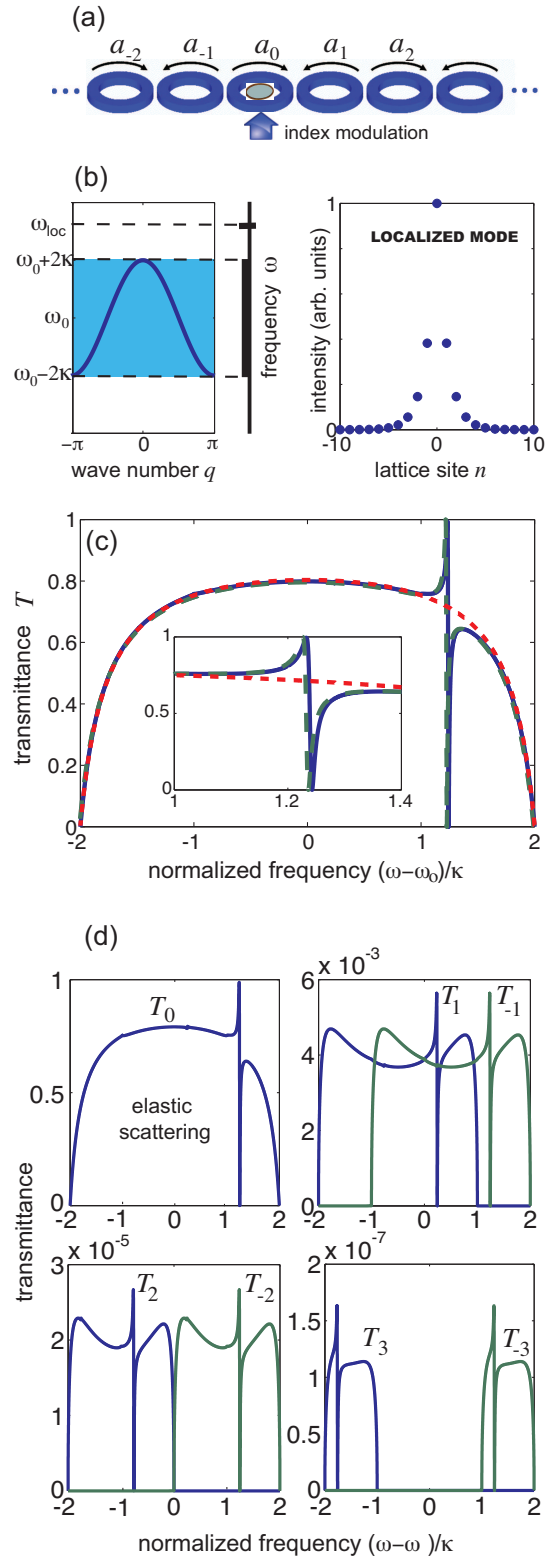}
\caption{(Color online) (a) Schematic of a CROW with dynamic modulation. (b) Left panel: energy spectrum of the CROW, comprising the continuous band $(\omega_0-2 \kappa, \omega_0+ 2 \kappa)$ of Bloch modes (dashed area) with dispersion relation $\omega(q)=\omega_0+2 \kappa \cos q$ (solid curve). For $\sigma>0$, an additional bound state does exist, at a frequency $\omega_{loc}$ above the continuous band. Right panel: intensity distribution of the bound mode for $\sigma / \kappa=1$. (c)Transmittance versus normalized frequency $(\omega(q)-\omega_0) / \kappa$ for $\sigma / \kappa=1$, $\Omega / \kappa=1$ and $\Gamma / \kappa=0.3$ (solid curve). The dotted curve is the transmittance of the static CROW ($\Gamma=0$), whereas the dashed curve is the transmittance as predicted by the approximate analytical relation (6). Inset: enlargement of the Fano resonance. (d) The four panels in the figure show the contribution to the total transmittance [solid curve in panel (c)] that arises from the elastic scattering channel ($T_0$) and from the six inelastic scattering channels ($T_{\pm1}$, $T_{\pm 2}$ and $T_{\pm 3}$).}
\end{figure*}
The transmittance $T_{st}(q)$ of the static CROW is simply given by $T_{st}(q)=|t_{st}(q)|^2$.
A typical behavior of $T_{st}$ versus the normalized frequency $(\omega(q)-\omega_0)/ \kappa$ is shown in Fig.1(c), dotted curve. Dynamic Fano resonances can be created by switching on the ac modulation term. To this aim, let us assume for the sake of definiteness $\sigma>0$, however a similar analysis holds for $\sigma<0$. The special case $\sigma=0$ behaves rather differently and it will be considered in the next section. For $\sigma>0$ the static CROW structure sustains a localized (bound) mode oscillating at the frequency $\omega_{loc}=\omega_0+ (4 \kappa^2 + \sigma^2)^{1/2}$ outside the transmission CROW band; see Fig.1(b). The bound mode is exponentially localized around the site $n=0$ with a localization length $1/ \mu$ defined by the relation $\sinh \mu= \sigma / (2 \kappa)$, namely one has  $|c_n|^2 = \exp(- 2 \mu |n|)$ [see the right panel in Fig.1(b)].
\begin{figure}
\includegraphics[scale=0.4]{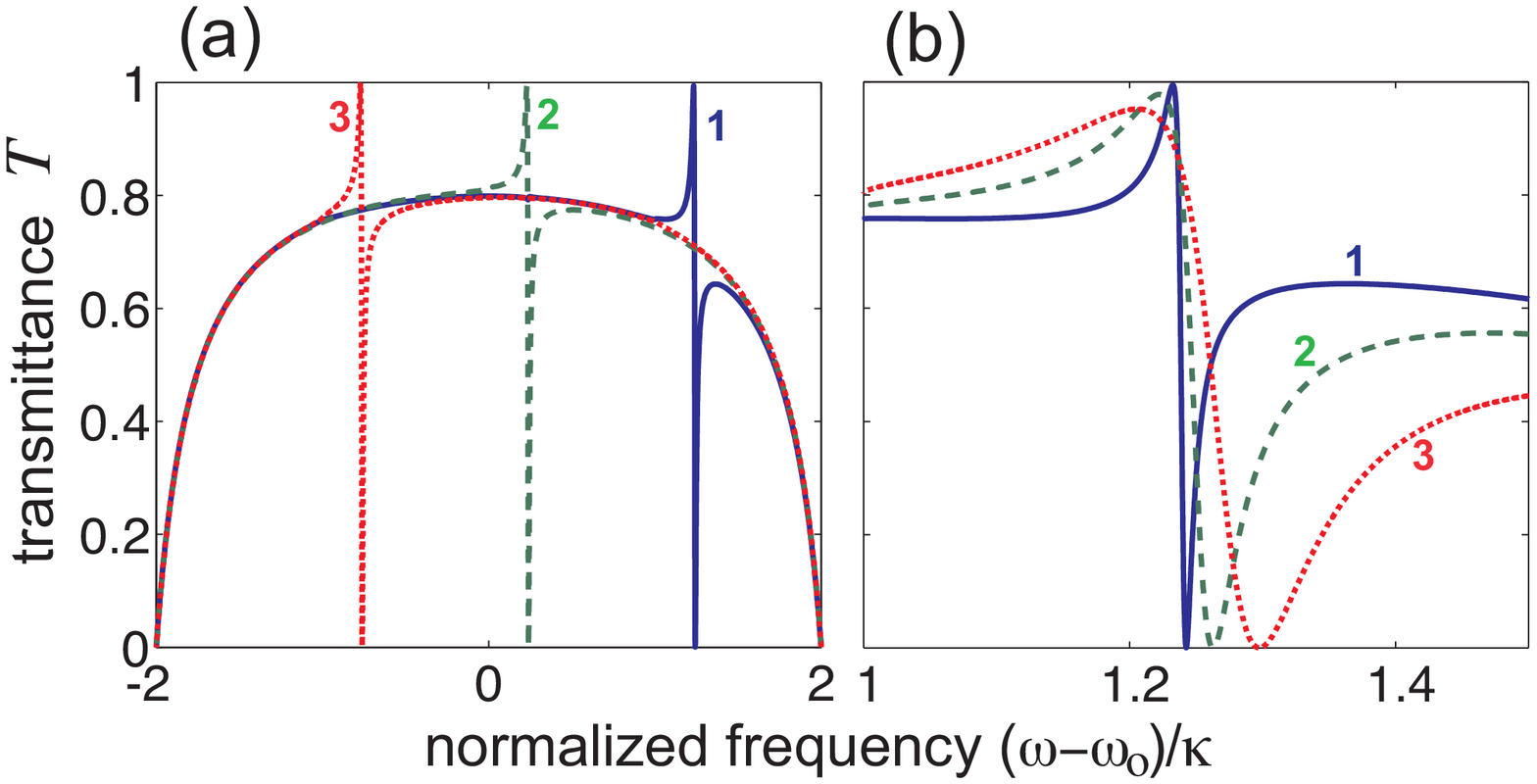}
\caption{(Color online) Fano lineshape engineering. (a) Transmittance for $\sigma/ \kappa=1$, $\Gamma / \kappa=0.3$ and for $\Omega/ \kappa=1$ (curve 1), $\Omega/ \kappa=2$ (curve 2), and $\Omega/ \kappa=3$ (curve 3). (b) Transmittance for $\Omega/ \kappa=1$, $\sigma/ \kappa=1$ and for $\Gamma/ \kappa=0.3$ (curve 1), $\Gamma / \kappa=0.6$ (curve 2), and $\Gamma / \kappa=0.9$ (curve 3).}
\end{figure}
To create a Fano resonance at the frequency $\omega_F$ inside the CROW transmission band, let us modulate the microcavity refractive index at the frequency $\Omega$ satisfying the condition $\Omega=\omega_{loc}-\omega_F$. In this case, a light photon at the frequency $\omega(q)$ close to $\omega_F$  incident onto the modulated microresonator (incident Floquet channel) can gain one energy quantum $ \hbar \Omega$ and drop into the
'bound' state. Similarly, photons in the 'bound' state can
loss  one energy quantum $\hbar \Omega$ and jump to the incident channel. Interference between direct and indirect (bound-state mediated) photon crossing creates
 a Fano-like resonance, as shown in Fig.1(c), solid curve. For parameter values chosen in the simulation, the total transmittance $T$ comes from the contribution of the scattering channel $T_0$ and of six inelastic channels $T_{\alpha}$, corresponding to $\alpha= \pm 1, \pm 2, \pm 3$ (for higher values of $|\alpha|$ the modes are evanescent at any incidence photon energy and do not contribute to the transmittance). The different (elastic and inelastic) contributions to the total transmittance are depicted in Fig.1(d). Note that the contribution from the inelastic 
 scattering channels is extremely small, i.e. the transmitted and reflected light basically has the same frequency than the one of the incident light. Formally, the onset of the Fano resonance can be explained by considering the small-modulation limit $\Gamma / \kappa \rightarrow 0$ and looking for a solution to Eqs.(5) as a power series $t_{\alpha}=t_{\alpha}^{(0)}+t_{\alpha}^{(1)}+t_{\alpha}^{(2)}...$, where the term $t_{\alpha}^{(k)}$ is of the order $ \sim (\Gamma / \kappa)^k$. At leading order ($k=0$) one recovers the static case, namely $t_{\alpha}^{(0)}(q)= \delta_{\alpha,0} t_{st}(q)$, whereas at first order one obtains
\begin{equation}
t_{\alpha}^{(1)}(q)= \left\{
\begin{array}{cc}
0 & \alpha \neq \pm 1 \\
-\frac{\Gamma}{2 \kappa} \frac{t_{st}(q)  \exp(iq)}{1-\exp(2iq_{\alpha})+(\sigma/ \kappa) \exp(iq_{\alpha})} & \alpha= \pm1
\end{array}
\right.
\end{equation}
While $t_{-1}^{(1)}(q)$ is a bounded function of $q$ and of order $\sim \Gamma / \kappa$, $t_{1}^{(1)}(q)$ shows a singularity at $q=q_F={\rm arc cos} [(\omega_{loc}-\Omega-\omega_0)/ 2 \kappa]$, which breaks the perturbative analysis. The singularity can be avoided by considering both $t_{0}$ and $t_{1}$ of the same order  $\sim 1$. In the two-channel approximation, one obtains at leading order
\begin{equation}
t_{0}^{(0)}(q) \simeq \frac{\exp(-2iq)-1}{1-\exp(2iq)+(\sigma / \kappa) \exp(iq)-\Sigma(q)}
\end{equation}
where we have set $\Sigma(q)=(\Gamma / 2 \kappa)^2 \exp[i(q+q_1)]/[1-\exp(2iq_1)+(\sigma/ \kappa) \exp (iq_1)]$. Note that $t_{0}^{(0)}(q)$ differs from the static value $t_{st}(q)$ because of the complex term $\Sigma(q)$ in the denominator of Eq.(8). Since near $q \sim q_F$ the Floquet channel $\alpha=1$ corresponds to an evanescent mode, at leading order the transmittance can be thus calculated as $T(q) \simeq |t_{0}^{(0)}|^2$, which well reproduces the exact shape obtained by exact numerical analysis of Eqs.(4) [see the dashed curve in Fig.1(c)]. Note that, around the frequency $\omega=\omega_F$, the transmittance shows a characteristic sharp and asymmetric profile, rapidly varying from zero to (almost) one. The frequency $\omega_F$ around which the Fano resonance appears can be tuned by changing the modulation frequency $\Omega$ according to the relation $\omega_F=\omega_{loc}-\Omega$, whereas the resonance width  is controlled by the modulation amplitude $\Gamma$; see Fig.2.
\section{Resonance overlap and electromagnetically-induced transparency}
The case of unbiased microresonator ($\sigma=0$) shows a different behavior. Here the static transmittance is unity [Eq.(6) with $\sigma=0$], and the structure does not sustain any localized mode. When the modulation is switched on, for $\Omega< 2 \kappa$ quite remarkably {\it two} resonances (rather than one) at the frequencies $\omega_0+2 \kappa -\Omega$ and $ \omega_0 -2 \kappa+ \Omega$ are created, symmetrically placed with respect to $\omega_0$; see Fig.3(a). In the weak modulation limit, this behavior can be explained by observing that, for $\sigma=0$, a singularity can appear in {\it either} $t^{(1)}_{\pm1}(q)$ [Eq.(7)] at the Bloch wave numbers $q$ such that $\omega(q)= \omega_0 \pm ( 2 \kappa- \Omega)$. The two resonances appear as asymmetric dips in the transmission spectrum [see Fig.3(a)].
As $\Omega \rightarrow 2 \kappa^-$, the two resonances interference and overlap at $\omega=\omega_0$, yielding a more complex resonance pattern comprising a narrow transmission dip with embedded an ultra-narrow resonance peak with unity transmittance at frequency $\omega=\omega_0$, see Fig.3(b). The ultra-narrow transmission peak created inside the dark dip and arising form resonance overlapping can be regarded as a kind of electromagnetically-induced transparency  (EIT)  effect. The entire dynamical process, i.e. creation and overlapping of resonances, is well described within a five-channel model by considering the amplitudes $t_{\alpha}$, $\alpha=0, \pm1, \pm2$ in Eqs.(5).
\begin{figure}
\includegraphics[scale=0.4]{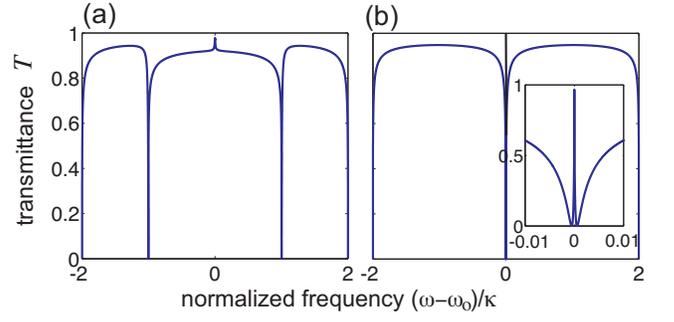}
\caption{(Color online) Spectral transmittance in the unbiased case ($\sigma=0)$ for $\Gamma / \kappa =0.8$ and for (a) $\Omega / \kappa=1$, and (b) $\Omega / \kappa=2$. In (b) the inset shows an enlargement of the transmittance around $\omega=\omega_0$, with a characteristic EIT-like spectrum arising from resonance overlap.}
\end{figure}
\begin{figure}
\includegraphics[scale=0.37]{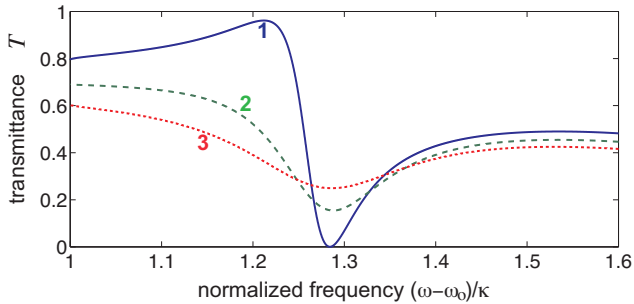}
\caption{(Color online) Effect of loss modulation on dynamic Fano resonances. Spectral transmittance for $\sigma / \kappa=1$, $\Gamma / \kappa =0.8$, $\Omega / \kappa=1$  and for increasing values of loss modulation parameter $\rho$: $\rho=0$ (curve 1), $\rho= 0.125$ (curve 2), and $\rho=0.25$ (curve 3).}
\end{figure}
\section{Experimental implementation and effects of cavity losses}
Let us briefly comment on the practical requirements for
the experimental implementation of dynamic Fano resonances. 
Ultrahigh-$Q$ coupled nanocavities can be realized in a two-dimensional triangular-lattice air-hole photonic
crystal slab \cite{r22bis}, and ultrafast dynamic
modulation of the refraction index can be achieved by carrier injection.
For CROW designed to operate at $\lambda_0 \sim 1560$ nm,
a typical value of the coupling rate is $\kappa \simeq 2 \pi \times 60$ GHz, according the experimental data of Ref.\cite{r22bis}. 
Hence the spectrum shown in Fig.1(c) corresponds to a dynamic Fano resonance with spectral width (distance between minimum and maximum of transmittance) of $\simeq 594$  MHz, created by refractive index modulation at frequency $\Omega=2 \pi \times 60$ GHz and modulation amplitude $\Gamma / \omega_0 \simeq 4.68 \times 10^{-5}$.
Assuming that the fractional change in
resonant frequency is directly related to the fractional change in
index of refraction $n$, i.e. $\Gamma / \omega_0 \simeq \delta n / n$, the estimated value of modulation is within the current technology limits of the maximum index change
($( \delta n / n )_{\max} \sim 10^{-4}$). Finally, let us briefly discuss the limitations introduced by cavity losses. There are basically two types of resonator losses that limit  the observation of Fano resonances: static cavity losses, and spurious time-periodic losses arising from index modulation of the resonator at site $n=0$. 
As far as static losses are concerned, let as assume a nanocavity $Q$-factor larger than $10^{6}$, which is achievable with current technology according to Ref. \cite{r22bis}; the static resonator losses simply limit the resolution of the spectrum to less than 200 MHz; the Fano resonance of Fig.1(c), corresponding to a spectral width of $\simeq 594$ MHz, should be therefore resolved. The other limitation that might prevent the observation of dynamic Fano resonances comes from carrier injection used to modulate the resonance frequency, which generally introduces spurious
dynamical modulation of cavity losses \cite{r36} at site $n=0$. We simulated the effect of cavity loss modulation by assuming $\Delta \omega_0(t)=\sigma+\Gamma \cos (\Omega t)-i \Gamma \rho [1-\cos(\Omega t)]$, where the dimensionless parameter $\rho$ accounts for the imaginary-to-real modulation strength induced by carrier injection \cite{r36}. Figure 4 shows an example of computed Fano resonances that takes into account modulation of resonator losses. The figure clearly shows that for moderate values of $\rho$ (e.g. $\rho>0.2$) a severe degradation of the Fano resonance occurs. Hence spurious dynamic modulation of cavity losses represents the most severe limitation for the observation of narrow Fano resonances; in particular carrier injection leading to modulation strengths $\rho$ of cavity loss larger than $ \sim 0.2$ should be avoided. 

\section{Conclusions}
Lineshape engineering, i.e. the possibility to control and tuning frequency and shape of the Fano resonance, is a functionality of key relevance. Conventionally, the Fano lineshape can be tuned over wide spectral ranges by carefully altering the geometry of a nanostructure, however  material or geometric engineering alone does not enable dynamical and fine lineshape control. In this work a novel dynamic method to create, shape and tuning Fano resonances in photonic structures has been theoretically proposed, which is based on dynamic modulation of the resonance frequency of a micro/nano cavity. Analytical results based on exact and perturbative Floquet analysis have been presented for the case of coupled resonator optical waveguide structures, however  the method can be extended to other photonic structures, such as a waveguide side coupled to a cavity \cite{r39}. 
Our results provide an important step toward the engineering and dynamical control of Fano resonances in integrated photonic structures that could be extended to disclose new physical  aspects and applications of Fano resonances. For example,  the use of dynamic modulation breaks time reversal symmetry and thus, with a judicious engineering of the photonic structure, could be potentially extended to realize non-reciprocal Fano-like resonances, i.e. different Fano line shapes for forward and backward propagation, a possibility that has been so far overlooked.

\end{document}